\documentclass[a4paper,11pt]{article}
\pdfoutput=1 

\usepackage{jheppub} 
\usepackage{slashed}
\usepackage[vcentermath]{youngtab}
\usepackage[T1]{fontenc} 

\newcommand{\beq}{\begin{eqnarray}}
\newcommand{\eeq}{\end{eqnarray}}

\newcommand{\bmp}{\noindent\begin{minipage}{16cm}}
\newcommand{\emp}{\end{minipage}\vskip 7mm} 

\newcommand{\SU}{\mbox{SU}}
\newcommand{\SO}{\mbox{SO}}
\newcommand{\SP}{\mbox{Sp}}
\newcommand{\UU}{\mbox{U}}

\definecolor{pumpkin}{rgb}{1.0, 0.4, 0.0}

\title{Electroweak precision tests of composite Higgs models}

\author[a]{Mads T. Frandsen}
\author[b]{and Martin Rosenlyst}

\affiliation[a]{CP$^3$-Origins, University of Southern Denmark, Campusvej 55, DK-5230 Odense M, Denmark} 
\affiliation[b]{Rudolf Peierls Centre for Theoretical Physics, University of Oxford, 1 Keble Road, Oxford OX1 3NP, United Kingdom\\}

\emailAdd{frandsen@cp3.sdu.dk}
\emailAdd{martin.jorgensen@physics.ox.ac.uk}

\abstract{We study constraints on Composite Higgs models with fermion partial compositeness from electroweak precision measurements, including the 2022 $W$-boson mass result from the CDF collaboration. We focus on models where the Composite Higgs sector arises from underlying four-dimensional strongly interacting gauge theories with fermions, and where the SM fermions obtain their mass via linear mixing terms between the fermions and the composite sector --- the so-called fermion partial compositeness scenario. In general, the Composite Higgs sector leads to a small and positive $S$ parameter, and a negative $T$ parameter, but the fermion partial compositeness sector results in an overall positive $T$ parameter in a large part of parameter space. We, therefore, find good agreement between the full composite models and the current electroweak precision measurement bounds on $S$ and $T$ from LEP and CDF, including the offset and correlation of $S,T$ with respect to the SM predictions. }

\begin{document} 
\maketitle
\flushbottom

\section{Introduction}

The fundamental origin of the Higgs sector of the Standard Model (SM) and the related naturalness and triviality problems are important  problems in particle physics. A dynamical origin of the Higgs sector and electroweak symmetry breaking (EWSB) is an appealing possible solution of these problems. In ``Composite Higgs'' (CH) models~\cite{Kaplan:1983fs,Dugan:1984hq}, the Higgs boson arises as a composite pseudo-Nambu-Goldstone boson (pNGB) from a spontaneously broken global symmetry.~\footnote{In general, composite Higgs models were first proposed in Refs.~\cite{Terazawa:1976xx,Terazawa:1979pj}, but in these models the Higgs boson is not identified as a light pNGB candidate. } It can therefore be naturally light with respect to the compositeness scale and other composite resonances in the model --- in agreement with current observations. 

In this paper, we will further assume the global symmetry arises from the chiral symmetries of fermions of an underlying confining gauge-fermion theory. The interactions between the SM sector and the new composite sector breaks the global symmetry group $ G $ explicitly and triggers dynamical EWSB yielding masses for the SM electroweak gauge bosons. 
We also assume that the SM fermions gain their masses through a linear mixing of the SM fermions with composite fermions of the new strongly interacting sector, the so-called fermion ``Partial Compositeness'' (PC) mechanism~\cite{Kaplan:1991dc}. Generating the SM fermion masses, including the heavy top quark mass, without violating  flavour changing neutral current (FCNCs) constraints is a well-known challenge in composite models~\cite{Cacciapaglia:2020kgq}. The PC mechanism provides a possibility to achieve this provided the composite fermionic operators that linearly coupled to the top quark have sufficiently large anomalous dimensions. 
The PC operators also contribute to the so-called oblique electroweak parameters $S$, $T$ and $U$~\cite{Peskin:1991sw,Barbieri:2004qk}, in particular to the $T$ parameter. In Ref.~\cite{Grojean:2013qca}, it is e.g. shown that the PC sector in the SO(5)/SO(4) CH model may contribute positively to the $ T $ parameter, which is also reviewed in details in Ref.~\cite{Panico:2015jxa}.
On the other hand, the composite Higgs sector itself contributes to a positive $S$ parameter. 

The contributions to the $S$ and $T$ parameters from CH models can be tested by the precise measurements of electroweak precision observables performed at  LEP~\cite{ALEPH:2013dgf}, Tevatron ~\cite{CDF:2022hxs} and LHC~\cite{ATLAS:2017rzl}. 
Recently, the CDF collaboration published an updated more precise measurement of the W boson mass, $ m_W^{\rm CDF}=80,433.5\pm 9.4 $~MeV~\cite{CDF:2022hxs}, that differs with $ 7\sigma $ significance from the SM prediction, $ m_W^{\rm SM}=80,357\pm 6 $~MeV~\cite{ParticleDataGroup:2020ssz}, and also from the LEP, D\O, CDF and ATLAS measurements: $ m_W^{\rm LEP}=80,385\pm 15 $~MeV from LEP~\cite{ALEPH:2010aa} and $ m_W^{\rm ATLAS}=80,370\pm 19 $~MeV from ATLAS~\cite{ATLAS:2017rzl}. 

By assuming $ U=0 $, the recent CDF measurement of the $W$ mass can be translated to best fit values for the  $ S $ and $ T $ parameters~\cite{Strumia:2022qkt,Lu:2022bgw}: \begin{equation}
\begin{aligned}
S=0.06\pm 0.08\,, \quad \quad T=0.15\pm 0.06 \label{eq: CDF results}
\end{aligned}
\end{equation} with a strong correlation $ \rho \approx 0.95 $. while the previous Particle Data Group (PDG) value before the CDF measurement was~\cite{ParticleDataGroup:2020ssz}\begin{equation}
\begin{aligned}
S=0.00\pm 0.07\,, \quad \quad T=0.05\pm 0.06 \label{eq: precision measurements results}
\end{aligned}
\end{equation} with the correlation $ \rho \approx 0.92 $. 



In this paper, we explore the parameter space of Composite Higgs models in light of these electroweak precision constraints. In particular, we show that the observed offset and correlation of $S$ and $T$, with respect to the SM  prediction, arises in a large part of the explored parameter space. I.e. in a large part of the viable parameter space, we find a positive  contribution to the $ S $ parameter from the composite higgs sector and a positive contribution to $T$ from the fermion PC sector --- in agreement with the combined  LEP and CDF precision measurements.~\footnote{In Ref.~\cite{Cacciapaglia:2022xih}, it is shown that this agreement is also naturally accommodated in other dynamical models such as the dilaton Higgs, the Technicolor and glueball Higgs. }



We focus on a minimal CH model realization with $ G/H=\SU(4)/\SP(4)\sim \SO(6)/\SO(5) $ \cite{Galloway:2010bp} as template~\footnote{The smaller $ \SO(5)\rightarrow \SO(4) $ coset is difficult to realize in a simple manner from a 4D gauge-fermion theory but arises more simply in the 5D holographic approach~\cite{Agashe:2004rs}.}, and provide a discussion of how our results generalize to non-minimal CH models. In the CH framework with extended global symmetries, the dark matter may be predicted as composite pNGB candidates, typically producing thermal DM~\cite{Ma:2015gra,Wu:2017iji,Cai:2018tet,Alanne:2018xli,Cacciapaglia:2019ixa} and in some models non-thermal candidates~\cite{Cai:2019cow,Cai:2020bhd,Cacciapaglia:2021aex,Rosenlyst:2021elz}. These models may be tested by the results in this paper. Furthermore, if the 2022 CDF results are interpreted in terms of Composite Higgs models they may offer a striking first window into origin of the SM fermion masses. It will be very interesting to further explore the implications for distinguishing between different posible fermion mass origins, in particular ``Extended Technicolor'' (ETC)~\cite{Dimopoulos:1979es} and the complementarity with other probes of these mass origins \cite{Alanne:2016rpe}.

\section{A minimal composite Higgs model}
The possible chiral symmetry breaking patterns in CH models based on an underlying four-dimensional gauge theory of strongly interacting fermions (hyper-fermions) are discussed in Refs.~\cite{Witten:1983tx,Kosower:1984aw}. Given $N_f$ Weyl spinors transforming in the representation $\mathcal{R}$  of the confining hypercolor (HC) group $ G_{\rm HC} $, the three possible vacuum cosets upon chiral symmetry breaking are~\cite{Peskin:1980gc}: $ \SU(N_f)/\SO(N_f) $ for real $\mathcal{R}$, $ \SU(N_f)/\SP(N_f) $ for pseudo-real $\mathcal{R}$ and $ \SU(N_f)\otimes \SU(N_f)\otimes \UU(1)/ \SU(N_f)\otimes \UU(1)$ for complex $\mathcal{R}$. The minimal CH cosets for these three types contain $N=5$ in the real case~\cite{Dugan:1984hq}, $N=4$ in both the pseudo-real~\cite{Galloway:2010bp} and the complex cases~\cite{Ma:2015gra}. In terms of pNGB spectrum, the pseudo-real case is the most minimal with only five composite pNGB states. Therefore, we consider concretely the minimal coset $ \SU(4)/\text{Sp}(4) $~\cite{Galloway:2010bp} providing both a Higgs candidate and custodial symmetry, where hyper-fermions and their representations under the electroweak gauge group and the HC gauge group $ G_{\rm HC} $ are listed in Table~\ref{tab: fermion content}. 

\begin{table}[t]
    \label{tab:su4}
    \begin{center}
	\begin{tabular}{cccc}
	    \hline
	    & $\vphantom{\frac{\frac12}{\frac12}}\quad G_{\mathrm{HC}}\quad$ & $\SU(2)_{\mathrm{W}}\quad$ 
	    & $\mathrm{U}(1)_{\mathrm{Y}}\quad$ \\
	    \hline
	    $\vphantom{\frac{\frac12}{\frac12}} (U_L,D_L)$	&   ${\tiny \yng(1)}$	&   ${\tiny \yng(1)}$	&   0\\    
	    $\vphantom{\frac{1}{\frac12}} \widetilde{U}_{\mathrm{L}}$	&   ${\tiny \yng(1)}$	&   1	&   $-1/2$\\    
	    $\vphantom{\frac{1}{\frac12}} \widetilde{D}_{\mathrm{L}}$	&   ${\tiny \yng(1)}$	&   1	&   $+1/2$ 
	    \\  
	    \hline
	\end{tabular}
    \end{center}
       \caption{The new fermion content and their representations under the gauge groups. 
   \label{tab: fermion content} }
\end{table}

\subsection{The underlying Lagrangian}

This minimal symmetry breaking pattern $ \SU(4)\rightarrow \SP(4) $ can thus be achieved if the hyper-fermions are in a pseudo-real representation of the HC gauge group: this can be minimally achieved for $ \rm G_{\rm HC} = \rm \SU(2)_{HC} $ or $ \rm \SP(2N)_{\rm HC} $ (or $ \SO(N)_{\rm HC} $) with the hyper-fermions in the fundamental (or spin) representation. The four Weyl hyper-fermions can be arranged into an $ \SU(4)_Q $ vector $ Q \equiv (U_L,D_L,\widetilde{U}_L,\widetilde{D}_L)^T $. In terms of this fourplet, the underlying gauge-fermion Lagrangian of the CH model can be written as \beq 
\mathcal{L}_{\rm CH}= Q^\dagger i \gamma^\mu D_\mu Q - \frac{1}{2} \left(Q^T M_Q Q +{\rm h.c.} \right)
	 + \mathcal{L}_{\rm PC}\,, 
\label{eq: Basic Lagrangian (UV)}
\eeq where the covariant derivative include the HC--gluons and the 
$\SU(2)_{L}$ and $ \UU(1)_{Y} $ gauge bosons. The mass term contributes to the correct vacuum alignment in between the EW unbroken vacuum and a Technicolor vacuum, and consists of two independent masses $ \overline{m}_{1,2} $ for the doublet and singlet hyper-fermions, $M_Q=\text{Diag}(i\overline{m}_1\sigma_2,-i\overline{m}_2\sigma_2)$, where $ \sigma_2 $ is the second Pauli matrix. 

The terms in $ \mathcal{L}_{\rm PC }$ in Eq.~(\ref{eq: Basic Lagrangian (UV)}) are interactions responsible for generating the masses and Composite Higgs-Yukawa couplings for the SM fermions in the condensed phase. The PC operators of $ \mathcal{L}_{\rm PC }$ require the extension of the model in Table~\ref{tab: fermion content} by a new species of fermions $\chi_t$, transforming under the two-index anti-symmetric representation of $ G_{\rm HC}=\SP(2N)_{\rm HC} $, and carrying appropriate quantum numbers under the SM gauge symmetry. For the top, it is enough to introduce a $\SU(2)_W$ vector-like pair with hypercharge $+2/3$ and transforming as a fundamental of the ordinary $ \SU(3)_C $ color gauge group (QCD). Models of this type were first proposed in Refs.~\cite{Barnard:2013zea,Ferretti:2013kya} and our model is an extension of the one in Ref.~\cite{Barnard:2013zea}. 

We  arrange the new QCD colored hyper-fermions into an $ \SU(6)_\chi $ vector, which will spontaneously break to $ \SO(6)_\chi $ upon the condensation. Thus, we study the minimal CH model with global symmetry breaking pattern \beq \label{eq: the chiral symmetry breaking}
\frac{\SU(4)_Q}{\SP(4)_Q}\otimes\frac{\SU(6)_\chi}{\SO(6)_\chi}\otimes \frac{\UU(1)_Q \otimes \UU(1)_\chi}{\varnothing}\,,
\eeq where one of the subgroups of $ \UU(1)_Q \otimes \UU(1)_\chi $ has an anomaly with the hypercolor symmetry group, $ G_{\rm HC} $, while the other one is an anomaly-free subgroup $ \UU(1)_\sigma $. Moreover, the QCD gauge group $ \SU(3)_C $ will be identified as a subgroup of the unbroken group $ \SO(6)_\chi $. This global symmetry breaking pattern can thus determine the possible hypercolor groups~\cite{Ferretti:2013kya}, which can only be $ G_{\rm HC}=\SP(2N)_{\rm HC} $ with $ 2\leq N \leq 18 $ (or $ G_{\rm HC}=\SO(N)_{\rm HC}$ with $ N=11,13 $ for $ Q $ and $ \chi $ transforming, respectively, in the spin and fundamental representations under $ G_{\rm HC} $). Therefore, the minimal choice is $ G_{\rm HC}=\SP(4)_{\rm HC} $. 


The four-fermion interactions that will generate the PC operators we consider are~\cite{Alanne:2018wtp}: 
\beq \label{eq: top PC-operators}
 \frac{\widetilde{y}_{L}}{\Lambda_t^2} q^{\alpha\dagger}_{L,3} (Q^\dagger P_q^\alpha Q^*\chi_{t}^\dagger) + \frac{\widetilde{y}_{R}}{\Lambda_t^2}t_{R}^{c\dagger} (Q^\dagger P_t Q^* \chi_{t}^\dagger)+\rm h.c.\,, 
\eeq  
where $ q_{L,3} $ and $ t_R $ are the third generation of the quark doublets and the top singlet, respectively, and $P_q$ and $P_t$ are spurions that project onto the appropriate  components in the fourplet $Q$. For both the left- and right-handed top, we choose the spurions to transform as the two-index anti-symmetric of the unbroken chiral symmetry subgroup $ \SU(4)_Q $ as  the minimal possible choice. Concretely, the spurions are given in matrix form in Ref.~\cite{Alanne:2018wtp}.

\subsection{The effective Lagrangian}
Upon the condensation of the hyper-fermions at a scale $ \Lambda_{\rm HC}\sim 4\pi f $, where $f$ is the decay constant of the composite pNGBs they can form an anti-symmetric and SM gauge invariant condensate of the form
\beq
 \langle Q^I_{\alpha,a}Q^J_{\beta,b}\rangle\epsilon^{\alpha\beta}\epsilon^{ab}\sim \Sigma_{Q0}^{IJ}=\left(\begin{array}{cc} i\sigma_2 & 0 \\ 0 & -i\sigma_2 \end{array}\right) \,,
\eeq
where $ \alpha,\beta $ are spinor indices, $ a,b $ are HC indices, and $ I,J $ are flavour indices of the hyper-fermions. This condensate visibly breaks $ \SU(4)_Q\rightarrow \SP(4)_Q $ via an anti-symmetric tensor~\cite{Galloway:2010bp}, resulting in five pNGBs, $ \pi_A $ with $ A=1,\dots,5 $, corresponding to the broken generators, $ X_A $, and ten unbroken generators $ T_{\overline{A}} $ with $ \overline{A}=1,\dots,10 $. Here the unbroken generators $ T_{\overline{A}}=\{ T_L^a, T_R^a, T_r^i\} $ with $ a=1,2,3 $ and $ i=1,\dots,4 $, in which $ T_{L,R}^a $ belong to the custodial symmetry subgroup $ \SO(4)_C\cong \SU(2)_L \otimes \SU(2)_R $ while $ T_r^i $ belong to the coset $ \SP(4)_Q/\SO(4)_C $. In the following, we identify $ T_L^a $ and $ T_R^3 $ as the generators for the $ \SU(2)_L $ and $ \UU(1)_Y $ gauge groups, respectively. The explicit matrix form of all these generators are listed in Ref.~\cite{Galloway:2010bp}. 

We parameterize the pNGBs as $ U_Q=\exp[i\pi_A X_A /f] $, where $ f $ is their decay constant. In the following, we identify the composite pNGB Higgs candidate as $ h\equiv \pi_4 $ and a pseudo-scalar $ \eta\equiv \pi_5 $, while the additional three are the Goldstones eaten by the $W^\pm$ and $Z$ gauge bosons. In the unitary gauge, we can write \begin{equation}
\begin{aligned}
 U_Q=\left(\begin{array}{cc} (c_Q+i\frac{\eta}{\pi_Q}s_Q)\mathbf{1}_2 & i\sigma_2\frac{h}{\pi_Q}s_Q \\ i\sigma_2\frac{h}{\pi_Q}s_Q & (c_Q-i\frac{\eta}{\pi_Q}s_Q)\mathbf{1}_2 \end{array}\right)\,, \label{eq: UQ matrix}
\end{aligned}
\end{equation} where $ c_Q \equiv \cos (\pi_Q/(2f)) $ and $ s_Q \equiv \sin (\pi_Q/(2f)) $ with $ \pi_Q\equiv \sqrt{h^2+\eta^2} $. Finally, the explicit form of the pNGB matrix with coset $ \SU(4)_Q/\text{Sp}(4)_Q $ is given by $ \Sigma_Q=U_Q^2 \Sigma_{Q0} $.

Below the compositeness scale $ \Lambda_{\rm HC} $, Eq.~\eqref{eq: Basic Lagrangian (UV)} is replaced by the effective Lagrangian:
\beq
    \label{eq: effLag}
    \mathcal{L}_\mathrm{eff}=\mathcal{L}_{\mathrm{kin}}-V_{\mathrm{eff}}\,.
\eeq 
Here $ \mathcal{L}_{\mathrm{kin}} $ is the usual leading order ($\mathcal{O} (p^2)$) chiral Lagrangian~\cite{Cacciapaglia:2020kgq},\beq
    \label{eq:effLag kin}
\mathcal{L}_{\mathrm{kin},p^2}=\frac{f^2}{4}\text{Tr}[d_\mu d^\mu]\,,
\eeq where the $ d $ and $ e $ symbols of this model are given by the Maurer-Cartan form \begin{equation}
\begin{aligned}
iU_Q^\dagger D_\mu U_Q \equiv d_\mu^A X^A + e_\mu^{\overline{A}} T^{\overline{A}}\equiv d_\mu + e_\mu\,, \label{eq: d and e symbols}
\end{aligned}
\end{equation} with the gauge covariant derivative \begin{equation}
\begin{aligned}
D_\mu = \partial_\mu - i g W_\mu^a T_L^a - i g^\prime B_\mu T_R^3\,.
\end{aligned}
\end{equation} Besides providing kinetic terms in Eq.~(\ref{eq:effLag kin}) and self-interactions for the pNGBs, it will induce masses for the EW gauge bosons and their couplings with the pNGBs (including the SM Higgs identified as $ h $), \begin{equation}
\begin{aligned} &m_W^2 = \frac{1}{4}g^2 f^2 \sin^2 \left(\frac{\langle h\rangle}{f}\right)\,, \quad\quad m_Z^2=m_W^2/c^2_{\theta_W}\,, \\
&g_{hWW}=\frac{1}{4} g^2 f\sin \left( \frac{2\langle h\rangle}{f}\right)=g_{  hWW}^{\rm SM}\cos \left(\frac{\langle h\rangle}{f}\right)\,,\quad\quad g_{ hZZ}=g_{ hWW}/c^2_{\theta_W}\,, \label{eq: WZ masses and SM VEV}
\end{aligned}
\end{equation} where $ \theta_W $ is the Weinberg angle, $ g $ is the weak $ \SU(2)_{ L} $ gauge coupling and \begin{equation}
\begin{aligned} 
\sin\theta \equiv \sin\left(\frac{\langle h\rangle}{f}\right) \equiv  \frac{v_{\rm SM}}{f}\,, \quad \quad v_{\rm SM}=246~\text{GeV}\,. \label{SM VEV}
\end{aligned}
\end{equation} The vacuum misalignment angle $ \theta $ parametrizes the deviations of the CH Higgs couplings to the EW gauge bosons with respect to the SM Higgs. These deviations are constrained by direct LHC measurements~\cite{deBlas:2018tjm} of this coupling which imply an upper bound of $ s_\theta \lesssim 0.3 $. EW precision measurements also impose an upper limit which has been found to be stricter $ s_\theta \lesssim 0.2 $~\cite{Cacciapaglia:2020kgq}. However, in this study, we find that the constraints on $ s_\theta $ from EW precision measurements are alleviated when we include all contributions from the composite fermion resonances in the CH models

In Section~\ref{sec:The effective potential and vacuum misalignment}, we write the effective potential $ V_{\rm eff} $ in Eq.~(\ref{eq: effLag}) to leading order, which receives contributions from the hyper-fermion masses $ \overline{m}_{1,2} $ in Eq.~(\ref{eq: Basic Lagrangian (UV)}), and from integrating out the composite fermion, vector and axial-vector resonances in the effective theory.  By minimizing this effective potential $ V_{\rm eff} $, the misalignment angle $ \theta $ can be determined. The physics of these composite resonances are discussed in the following two sections.


\section{The effective theory of the spin-1 resonances}
\label{sec: The effective theory of the vector and axial-vector resonances}

The composite spin-1 resonances, analogous to the $ \rho $- and $a_1$-mesons in the QCD can be organized into a $ \textbf{10} $-plet $ V_\mu =V_\mu^{\overline{A}} T^{\overline{A}} $ of the unbroken $ \SP(4)_Q $ and a $ \textbf{5} $-plet  $ A_\mu = A_\mu^A X^A $  in $ \SU(4)_Q/\SP(4)_Q $  with $ \overline{A}=1,\dots,10 $ and $ A=1,\dots,5 $. Here $ T^{\overline{A}} $ and $ X^A $ are, respectively, the ten unbroken generators of $ \SP(4)_Q $ and the five broken generators in $ \SU(4)_Q/\SP(4)_Q $. Under the decomposition $ \SP(4)_Q\rightarrow \SU(2)_L \otimes \UU(1)_Y $, these objects decompose into \begin{equation}
\begin{aligned}
&\left(\begin{array}{l}\textbf{10} \rightarrow \textbf{3}_{0}\oplus \textbf{1}_{1}\oplus \textbf{1}_{0}\oplus \textbf{1}_{-1}\oplus \textbf{2}_{1/2}\oplus \textbf{2}_{-1/2}\\ V \rightarrow V_L \oplus V_R^+ \oplus V_R^0 \oplus V_R^- \oplus V_D \oplus \widetilde{V}_D  \end{array}\right)\,, \quad \left(\begin{array}{l}\textbf{5} \rightarrow \textbf{2}_{1/2}\oplus \textbf{2}_{-1/2}\oplus \textbf{1}_{0}\\ A \rightarrow A_D \oplus \widetilde{A}_D \oplus A_S  \end{array}\right)\,,
\end{aligned}
\label{Eq:vecs}
\end{equation} where $ \widetilde{V}_D = i\sigma_2 V_D^* $ and $ \widetilde{A}_D=i\sigma_2 A_D^* $.

In the following, we will use the hidden gauge symmetry formalism to describe the spin-1 resonances, as done in Ref.~\cite{BuarqueFranzosi:2016ooy}. We therefore formally extend the global symmetry group $ \SU(4)_Q $  to $ \SU(4)_1\otimes \SU(4)_2 $. We embed the elementary SM gauge bosons in $ \SU(4)_1 $ and the heavy composite spin-1 resonances in $ \SU(4)_2 $. In the effective Lagrangian, the $ \SU(4)_i $ ($ i=1,2 $) symmetries are spontaneously broken to $ \SP(4)_i $ via the introduction of two nonlinear representations $ U_i $ containing five pNGBs each (similar to the pNGB matrix $ U_Q $ in Eq.~(\ref{eq: UQ matrix})), which are given by \begin{equation}
\begin{aligned}
U_1=\exp\left(\frac{i}{f_1}\sum_{A=1}^5 \pi_1^A X^A\right)\,, \quad \quad U_2=\exp\left(\frac{i}{f_2}\sum_{A=1}^5 \pi_2^A X^A\right)\,.
\label{eq:}
\end{aligned}
\end{equation} They transform as $ U_i \rightarrow g_i U_i h(g_i,\pi_i)^\dagger $, where $ g_i $ and $ h(g_i,\pi_i) $ are group elements of $ \SU(4)_i $ and $ \SP(4)_i $, respectively.  Furthermore, the breaking of $ \SP(4)_1\otimes \SP(4)_2 $ into the chiral subgroup $ \SP(4)_Q $ is described by the field \begin{equation}
\begin{aligned}
K=\exp\left(\frac{i}{f_K}\sum_{\overline{A}=1}^{10}k^{\overline{A}} T^{\overline{A}}\right)\,,
\label{eq: K fields}
\end{aligned}
\end{equation} containing ten pNGBs corresponding to the generators of $ \SP(4) $ and transforming like $ K\rightarrow h(g_1,\pi_1) K h(g_2,\pi_2)^\dagger $. 

To write down the effective Lagrangian for the spin-1 resonances and $ K $, we define the gauged Maurer-Cartan one-forms as \begin{equation}
\begin{aligned}
\Omega_{i,\mu}&\equiv iU_i^\dagger D_\mu U_i \\ &\equiv d_{i,\mu}^A X^A + e_{i,\mu}^{\overline{A}} T^{\overline{A}}\equiv d_{i,\mu} + e_{i,\mu} \,, 
\label{eq: d and e symbols vec res}
\end{aligned}
\end{equation} where the covariant derivatives are given by 
\begin{equation}
\begin{aligned}
&D_\mu U_1 = \left(\partial_\mu -i g W_\mu^a T_L^a - i g^{\prime} B_\mu T_R^3\right)U_1\,, \\
&D_\mu U_2 = \left(\partial_\mu -i \widetilde{g} V_\mu^{\overline{A}}T^{\overline{A}} - i \widetilde{g} A_\mu^A X^A \right)U_2\,,
\label{eq:}
\end{aligned}
\end{equation} such that the heavy spin-1 vectors are formaly introduced like gauge fields similarly to the SM gauge fields. The covariant derivative of $ K $ has the form \begin{equation}
\begin{aligned}
D_\mu K =\partial_\mu K - i e_{1,\mu} K + iK e_{2,\mu}\,.
\label{eq:}
\end{aligned}
\end{equation} Here the $ d $ and $ e $ symbols transform under $ \SU(4)_i $ as follows \begin{equation}
\begin{aligned}
&d_{i,\mu} \rightarrow h(g_i,\pi_i) d_{i,\mu} h(g_i,\pi_i)^\dagger\,,  
\quad \quad e_{i,\mu} \rightarrow h(g_i,\pi_i) (e_{i,\mu}+i\partial_\mu)h(g_i,\pi_i)^\dagger\,.
\label{eq:}
\end{aligned}
\end{equation} Thus, the effective Lagrangian of the spin-1 resonances is given by \begin{equation}
\begin{aligned}
&\mathcal{L}_{\rm gauge} = -\frac{1}{2g^2} \text{Tr}[W_{\mu\nu}W^{\mu\nu}]-\frac{1}{2g^{\prime 2}}\text{Tr}[B_{\mu\nu}B^{\mu\nu}] - \frac{1}{2\widetilde{g}} \text{Tr}[V_{\mu\nu}V^{\mu\nu}] - \frac{1}{2\widetilde{g}} \text{Tr}[A_{\mu\nu}A^{\mu\nu}] \\ &\quad+ \frac{1}{2} f_1^2 \text{Tr}[d_{1,\mu}d^{\mu}_1]+ \frac{1}{2} f_{2}^2 \text{Tr}[d_{2,\mu}d^{\mu}_{2}]+r f_{2}^2 \text{Tr}[d_{1,\mu} K d^{\mu}_{2}K^\dagger]+ \frac{1}{2} f_K^2 \text{Tr}[D_\mu K D^\mu K^\dagger]\,,
\label{eq: VA lag}
\end{aligned}
\end{equation} where  \begin{equation}
\begin{aligned}
&W_{\mu\nu}=\partial_\mu W_\nu - \partial_\nu W_\mu -i g[W_\mu, W_\nu]\,,\quad \quad 
B_{\mu\nu}=\partial_\mu B_\nu - \partial_\nu B_\mu\,, \\ 
&V_{\mu\nu}=\partial_\mu V_\nu - \partial_\nu V_\mu -i \widetilde{g}[V_\mu, V_\nu]\,,\phantom{...} \quad \quad \quad 
A_{\mu\nu}=\partial_\mu A_\nu - \partial_\nu A_\mu -i \widetilde{g}[A_\mu, A_\nu]
\label{eq:}
\end{aligned}
\end{equation} with $ W_\mu = W_\mu^a T^a_L $. We have omitted terms containint a scalar singlet $ \sigma $ resonance, included in Ref.~\cite{BuarqueFranzosi:2016ooy}, as we here assume this resonance is heavy.

In terms of the above Lagrangian parameters, we define the mass parameters \begin{equation}
\begin{aligned}
m_V\equiv \frac{\widetilde{g}f_K}{\sqrt{2}}, \quad \quad m_A\equiv \frac{\widetilde{g}f_1}{\sqrt{2}}\,.  
\label{eq: mV and mA mass parameters}
\end{aligned}
\end{equation}  The masses of $ V_D $ and $ \widetilde{V}_D $ are $ m_V $, while the masses of $ A_S $ and the axial combination of $ A_D $ and $ \widetilde{A}_D $ are $ m_A $ as these vector resonances, given in Eq.~(\ref{Eq:vecs}), do not mix with the SM vector states. The explicit mass expressions of the additional gauge bosons are given in Ref.~\cite{BuarqueFranzosi:2016ooy}.


After integrating out the heavy resonances in Eq.~(\ref{eq: VA lag}), we obtain under the unitary gauge the Lagrangian \begin{equation}
\begin{aligned}
\mathcal{L}_{\rm eff}^{\rm gauge}=&\frac{1}{2}P_T^{\mu\nu} \bigg[\left(-p^2 + \frac{g^{\prime 2}}{g^2}\Pi_0(p)\right)B_\mu B_\nu + (-p^2+\Pi_0(p))W_\mu^a W_\nu^a  \\ & + \frac{\Pi_1(p)}{4}\frac{h^2}{f^2}\left(W_\mu^1 W_\nu^1+W_\mu^2 W_\nu^2+ \left(W_\mu^3-\frac{g^\prime}{g}B_\mu\right)\left(W_\nu^3-\frac{g^\prime}{g}B_\nu\right)\right)\bigg]\,,
\label{eq: gauge eff lag}
\end{aligned}
\end{equation} where the transverse and longitudinal projection operators are defined as \begin{equation}
\begin{aligned}
P_T^{\mu\nu}=g^{\mu\nu}-\frac{p^\mu p^\nu}{p^2}\,, \quad P_L^{\mu\nu}=\frac{p^\mu p^\nu }{p^2}\,,
\label{eq: form factors}
\end{aligned}
\end{equation} and the form factor are given by the Weinberg sum rules~\cite{Bian:2019kmg}  \begin{equation}
\begin{aligned}
&\Pi_0(p)= \frac{g^2 p^2 f_1^2}{p^2+m_V^2}\,, \quad \quad \Pi_1(p)= \frac{g^2 f^2 m_V^2 m_A^2}{(p^2+m_V^2)(p^2+m_A^2)}\,,
\label{eq: form factors}
\end{aligned}
\end{equation} where from the definition of the Fermi decay constant it is obtained that $ f=\sqrt{f_1^2-r^2 f_2^2} $~\cite{BuarqueFranzosi:2016ooy}. It will be shown in Section~\ref{sec:The effective potential and vacuum misalignment} that at one-loop level this Lagrangian contributes to the effective Higgs potential. 

\section{The effective theory of the fermion resonances}
\label{sec: The effective theory of the fermion resonances}

In this section, we introduce the effective theory of the fermion resonances, which are important for the generation of the masses and Composite Higgs-Yukawa couplings of the SM fermions. The composite fermion resonances also contribute to the effective Higgs potential in Eq.~(\ref{eq: eff pot total}) and affect the electroweak precision observables. 

We introduce a Dirac $ \textbf{6} $-plet of fermionic resonances $ \Psi $ transforming in the antisymmetric representation of $ \SU(4)_2 $ while the composite spin-1 resonances transform in the ajoint. The fermionc resonances will mix with the elementary doublet $ q_{L,3}=(t_L,b_L)^T $ and the singlet $ t_R $ to provide the top mass. 
A new $ \text{U}(1)_X $ gauge symmetry is introduced to provide the correct hypercharges, embedded in the unbroken $ \SO(6)_\chi $ in Eq.~(\ref{eq: the chiral symmetry breaking}). The SM hypercharge is defined as $ Y=T_{R,1}^3+T_{R,2}^3+X $, where $ T_{R,i}^3 $ are the diagonal generators of the subgroups $ \SU(2)_{R,i} $ in $ \SP(4)_i $. The SM $ \SU(2)_L $ is the vectorial combination of the $ \SU(2)_{L,i} $ subgroups in $ \SP(4)_i $. 

Now, we consider the $ \textbf{1} $ and $ \textbf{5} $ representations of the unbroken $ \text{Sp}(4)_2 $ in $ \SU(4)_2 $. 
Under the decomposition $ \text{Sp}(4)_2\otimes \text{U}(1)_X \rightarrow \SU(2)_L\otimes \text{U}(1)_Y $, we get the $\SU(2)_L$ singlet $ \Psi_1 $ (denoted $ T_1 $) with $ \mathbf{1}_{2/3} $ and the fiveplet \begin{equation}
\begin{aligned}
&\left(\begin{array}{l}\textbf{5}_{2/3}\rightarrow \textbf{2}_{7/6}\oplus \textbf{2}_{1/6}\oplus \textbf{1}_{2/3}\\ \Psi_5 \rightarrow Q_X \oplus Q_5\oplus T_5 \end{array}\right)\,, \label{eq: the fiveplet}
\end{aligned}
\end{equation} where $ Q_X=(X_{5/3},X_{2/3}) $ and $ Q_5=(T,B) $. Their explicit embeddings in representation $ \textbf{6} $ of $ \SU(4)_2 $ are~\cite{Dong:2020eqy} \begin{equation}
\begin{aligned}
&\Psi_{5L}=\frac{1}{2}\left(\begin{array}{cc} T_{5L} i\sigma_2 &\sqrt{2}\widetilde{Q}_L \\ -\sqrt{2}\widetilde{Q}_L^T &  T_{5L} i\sigma_2\end{array}\right)\,, \quad \quad \widetilde{Q}_L=\left(\begin{array}{cc} T_L &X_{5/3, L} \\ B_L &  X_{2/3,L}\end{array}\right)\,, \\
&\Psi_{5R}^c=\frac{1}{2}\left(\begin{array}{cc} T_{5R}^c i\sigma_2 &\sqrt{2}\widetilde{Q}_R^c \\ -\sqrt{2}\widetilde{Q}_R^{cT} &  T_{5R}^c i\sigma_2\end{array}\right)\,, \quad \quad \widetilde{Q}_R^c=\left(\begin{array}{cc} -X_{2/3,R}^c & B_R^c \\ X_{5/3,R}^c &  -T_R^c\end{array}\right)\,, \\
&\Psi_{1L}=\frac{1}{2}\left(\begin{array}{cc} T_{1L} i\sigma_2 &0 \\ 0 & -T_{1L} i\sigma_2\end{array}\right)\,, \quad\quad \Psi_{1R}^c=\frac{1}{2}\left(\begin{array}{cc} T_{1R}^c i\sigma_2 &0 \\ 0 & -T_{1R}^c i\sigma_2\end{array}\right)\,. \label{eq: explicit embeddings top partners}
\end{aligned}
\end{equation}


In the following, we focus on the case in which both the left- and right-handed top quarks mix with top-partners in the $ \mathbf{6} $ (two-index anti-symmetric) representation of $ \SU(4)_2 $, where the wave functions of these top-partners and their quantum number under $ \SU(4)_2\times \SU(6)_\chi $ in Eq.~(\ref{eq: the chiral symmetry breaking}) have the form $ QQ\chi\in (\mathbf{6},\mathbf{6}) $ as in the parentheses in Eq.~(\ref{eq: top PC-operators}). The left-handed fermionic operators of the form $ QQ\chi $ can be decomposed under the unbroken subgroup $ \text{Sp}(4)_2\times \SU(3)_C\times \text{U}(1)_X $ as \begin{equation}
\begin{aligned}
(\mathbf{6},\mathbf{6})&=(\mathbf{5},\mathbf{3},2/3)+(\mathbf{5},\overline{\mathbf{3}},-2/3)+(\mathbf{1},\mathbf{3},2/3)+(\mathbf{1},\overline{\mathbf{3}},-2/3)\\ &\equiv \Psi_{5L}+\Psi_{5R}^c+\Psi_{1L}+\Psi_{1R}^c\,,
\end{aligned}
\end{equation} where the explicit matrices of the composite resonances $ \Psi_{5L,5R,1L,1R} $ are given in Eq.~(\ref{eq: explicit embeddings top partners}). In order to describe the linear fermion mixing, the top quark doublet and singlet, $ q_L $ and $ t_R $ are embedded as a $ \textbf{6} $ of $ \SU(4)_Q $ as \begin{equation}
\begin{aligned}
&\Psi_{q_L}=\frac{1}{\sqrt{2}}\left(\begin{array}{cc}0 & Q_{q_L} \\ -Q_{q_L}^T & 0 \end{array}\right)\,, \quad \quad  Q_{q_L}=\left(\begin{array}{cc}t_L & 0 \\ b_L & 0 \end{array}\right)\,, \quad \quad \Psi_{t_R}^c= \frac{t_R^c}{2}\left(\begin{array}{cc} -i\sigma_2 & 0 \\ 0 & i\sigma_2 \end{array}\right)\,.
\end{aligned}
\end{equation} 

To leading order, the effective Lagrangian for the elementary and composite fermions can then be decomposed in three parts
 \begin{equation}
\begin{aligned}
\mathcal{L}_{\rm ferm}=\mathcal{L}_{\rm elem}+\mathcal{L}_{\rm comp}+\mathcal{L}_{\rm mix}\,. \label{eq: fermion resonances lag}
\end{aligned} 
\end{equation} The Lagrangian for the elementary fermions is \begin{equation}
\begin{aligned}
\mathcal{L}_{\rm elem}=& i \overline{q}_{L,3} \slashed{D} q_{L,3} + i \overline{t}_R \slashed{D}t_R \,,
\end{aligned} 
\end{equation} where the covariant derivatives is given by \begin{equation}
\begin{aligned}
D_\mu q_{L,3} = &\left(\partial_\mu - ig W_\mu^i \frac{\sigma_i}{2}-i\frac{1}{6}g^\prime B_\mu -ig_S G_\mu\right)q_{L,3}\,, \\
D_\mu t_L = &\left(\partial_\mu -i\frac{2}{3}g^\prime B_\mu -ig_S G_\mu\right)t_L
\,.
\end{aligned} 
\end{equation} The Lagrangian for the composite fermions containing their gauge-kinetic, mass and interaction terms between the fiveplet and singlet can be written as~\cite{Panico:2015jxa,Dong:2020eqy} \begin{equation}
\begin{aligned}
\mathcal{L}_{\rm comp}=& i \text{Tr}[\Psi_{5L}^\dagger \overline{\sigma}^\mu D_\mu \Psi_{5L}]+i \text{Tr}[(\Psi_{5R}^c)^\dagger \sigma^\mu D_\mu \Psi_{5R}^c]\\&+i \text{Tr}[\Psi_{1L}^\dagger \overline{\sigma}^\mu D_\mu \Psi_{1L}]+i \text{Tr}[(\Psi_{1R}^c)^\dagger \sigma^\mu D_\mu \Psi_{1R}^c]\\&-(m_5\text{Tr}[\Psi_{5L}\Sigma_{Q0}\Psi_{5R}^c\Sigma_{Q0}]+m_1\text{Tr}[\Psi_{1L}\Sigma_{Q0}\Psi_{1R}^c\Sigma_{Q0}]+\text{h.c.})\\&-(ic_L\text{Tr}[\Psi_{5L}^\dagger \overline{\sigma}^\mu d_{2,\mu} \Psi_{1L}]+ic_R\text{Tr}[\Psi_{5R}^\dagger \sigma^\mu d_{2,\mu} \Psi_{1R}]+\text{h.c.})
\,, \label{eq: comp lagrangian}
\end{aligned} 
\end{equation} where the covariant derivatives are given by \begin{equation}
\begin{aligned}
D_\mu \Psi_5 =&\left(\partial_\mu - i\frac{2}{3}g^\prime B_\mu T_R^3 -ie_{2,\mu} - ig_S G_\mu \right)\Psi_5\,, \\
D_\mu \Psi_1 =&\left(\partial_\mu -i \frac{2}{3}g^\prime B_\mu  T_R^3  - ig_S G_\mu  \right)\Psi_1\,,
\end{aligned} 
\end{equation} and the $ d $ and $ e $ symbols are defined in Eq.~(\ref{eq: d and e symbols vec res}) which contain the heavy spin-1 resonances. Finally, the mixing terms between the SM fermions, composite pNGBs and fermion resonances invariant under $ \SU(4)_Q $ are given by~\cite{Dong:2020eqy} \begin{equation}
\begin{aligned}
\mathcal{L}_{\rm mix}=&-y_{L5}f\text{Tr}[\Psi_{q_L}U_Q\Psi_{5R}^c U_Q^T]-y_{L1}f\text{Tr}[\Psi_{q_L}U_Q\Psi_{1R}^c U_Q^T]\\&-y_{R5}f\text{Tr}[\Psi_{t_R}^c U_Q\Psi_{5L}U_Q^T]-y_{R1}f\text{Tr}[\Psi_{t_R}^c U_Q\Psi_{1L} U_Q^T]+\text{h.c.}\,.
\end{aligned}
\end{equation}


From the effective Lagrangian $ \mathcal{L}_{\rm ferm} $ in Eq.~(\ref{eq: fermion resonances lag}), the mass matrix of the top quark and the top-partners with $ +2/3 $ charge has the form
\begin{equation}
\begin{aligned}
\left(\begin{array}{c} \overline{t}_L \\ \overline{T}_{1L} \\ \overline{T}_{L} \\ \overline{X}_{2/3L} \\ \overline{T}_{5L} \end{array}\right)^T \left(\begin{array}{ccccc} 0 & \frac{y_{L1}f}{\sqrt{2}}s_\theta & \frac{y_{L5}f}{2}(1+c_\theta) & \frac{y_{L5}f}{2}(1-c_\theta)& 0\\ y_{R1}fc_\theta & -m_1 & 0 & 0 & 0 \\ -\frac{y_{R5}f}{\sqrt{2}}s_\theta & 0 & -m_5 & 0 & 0 \\ \frac{y_{R5}f}{\sqrt{2}}s_\theta & 0 & 0 & -m_5 & 0 \\ 0 & 0 & 0 & 0 & -m_5 \end{array}\right) \left(\begin{array}{c} t_R \\ T_{1R} \\ T_{R} \\ X_{2/3R} \\ T_{5R} \end{array}\right)\,. \label{eq: top mass matrix}
\end{aligned}
\end{equation} By diagonalizing this mass matrix, the top mass is easily obtained \begin{equation}
\begin{aligned}
m_{\rm top}\simeq\frac{\vert y_{L1}y_{R1}m_5-y_{L5}y_{R5}m_1\vert f}{\sqrt{m_1^2 + y_{R1}^2f^2}\sqrt{m_5^2 + y_{L5}^2f^2}}\frac{v_{\rm EW}}{\sqrt{2}}\,,
\label{eq: top mass}
\end{aligned}
\end{equation} while the masses of top-partners with $ +2/3 $ charge (in the limit $ c_\theta \sim 1 $) are given by \begin{equation}
\begin{aligned}
m_{T_1}\simeq\sqrt{m_1^2+y_{R1}^2 f^2}\,,  \quad m_{T}\simeq\sqrt{m_5^2+y_{L5}^2 f^2}\,,  \quad m_{X_{2/3}}\simeq m_5\,, \quad m_{T_5}=m_5\,.
\label{eq: fermion resonance masses}
\end{aligned}
\end{equation} For $ m_{T}\gg m_{T_1} $, the top mass is approximately given by\begin{equation}
\begin{aligned}
m_{\rm top}\simeq\frac{y_{L1}y_{R1}f}{\sqrt{m_1^2 + y_{R1}^2f^2}}\frac{v_{\rm EW}}{\sqrt{2}} \simeq \frac{y_{L1}y_{R1}f}{m_{T_1}}\frac{v_{\rm EW}}{\sqrt{2}}\,.
\label{eq: top mass approx}
\end{aligned}
\end{equation} 

Finally, after integrating out the heavy top-partners in Eq.~(\ref{eq: fermion resonances lag}), we get the effective Lagrangian \begin{equation}
\begin{aligned}
\mathcal{L}_{\rm eff}^{\rm top}=&\Pi_0^q (p)\text{Tr}\left[\overline{\Psi}_{q_L}\slashed{p}\Psi_{q_L}\right]+\Pi_0^t (p)\text{Tr}\left[\overline{\Psi}_{t_R}^c \slashed{p}\Psi_{t_R}^c\right]\\ &+M_1^t (p)\text{Tr}\left[\Psi_{q_L}\Sigma_Q^*\Psi_{t_R}^c \Sigma_Q^*\right]+\text{h.c.}\,,
\label{eq: top eff lag}
\end{aligned}
\end{equation} where the form factors are given by the Weinberg sum rules~\cite{Bian:2019kmg} \begin{equation}
\begin{aligned}
&\Pi_0^q(p)=1+\frac{\vert y_{L5}\vert^2 f^2}{p^2+m_5^2}\,,\quad \Pi_0^t(p)=1+\frac{\vert y_{R5}\vert^2 f^2}{p^2+m_5^2}\,, \quad \quad \quad \quad  \\ &
M_1^t(p)=\frac{y_{L1}y_{R1}^* f^2 m_1}{p^2+m_1^2}-\frac{y_{L5}y_{R5}^* f^2 m_5}{p^2+m_5^2}\,.
\label{eq: form factors}
\end{aligned}
\end{equation} In the following section, we show that at one-loop level this Lagrangian contributes to the effective Higgs potential. 

\section{The effective potential and vacuum misalignment}
\label{sec:The effective potential and vacuum misalignment}

The vacuum alignment angle $\theta$ is controlled by the effective potential  $V_{\mathrm{eff}}$, which receives contributions from the vector-like masses of the hyper-fermions, the SM fermion couplings to the strong sector and the EW gauge interactions. At leading order, each source of symmetry breaking contributes independently to the effective potential:
\beq
V_{\text{eff}}&=& V_{\text{gauge}}+V_{\text{top}}+V_\text{m}+\dotsc\,, \label{eq: eff pot total}
\eeq 
where the dots are left to indicate the presence of mixed terms at higher orders, or the effect of additional UV operators.

From the effective Lagrangian in Eq.~(\ref{eq: gauge eff lag}), at one-loop level of the gauge interactions we can write the Coleman-Weinberg potential  \begin{equation}
\begin{aligned}  
V_{\rm gauge}= & \frac{6}{2} \int \frac{d^4 p}{(2\pi)^4}\log \left(1+\frac{\Pi_1}{4\Pi_W}\sin^2\left(\frac{\pi_Q}{f}\right)\right) \\ &+ \frac{3}{2}\int \frac{d^4 p}{(2\pi)^4}\log \left[1+\left(\frac{g^{\prime 2}}{g^2}\frac{\Pi_1}{4\Pi_B}+\frac{\Pi_1}{4\Pi_W}\right)\sin^2\left(\frac{\pi_Q}{f}\right)\right]\,,
\end{aligned}
\end{equation} where $ \Pi_W=p^2 + \Pi_0 $ and $ \Pi_B =p^2 + (g^\prime/g)^2 \Pi_0 $. To leading orders, the Higgs potential from the EW gauge boson loops is\begin{equation}
\begin{aligned}  
V_{\rm gauge}^0 = \alpha_g s_\theta^2+\beta_g s_\theta^4+\mathcal{O}(s_\theta^6)\,,
\end{aligned}
\end{equation} where \begin{equation}
\begin{aligned}  
\alpha_g \equiv &  \frac{3}{8}\int \frac{d^4 p}{(2\pi)^4} \left[\frac{3}{\Pi_W}+\frac{g^{\prime 2}}{g^2}\frac{1}{\Pi_B}\right]\Pi_1\,, \\ 
\beta_g \equiv & -\frac{3}{64}\int \frac{d^4 p}{(2\pi)^4} \left[2\left(\frac{1}{\Pi_W}\right)^2+\left(\frac{g^{\prime2}}{g^2}\frac{1}{\Pi_B}+\frac{1}{\Pi_W}\right)^2\right]\Pi_1^2\,.
\end{aligned}
\end{equation}

From the top quark effective Lagrangian in Eq.~(\ref{eq: top eff lag}), the Higgs potential $ V_{\rm top} $ at one-loop level in Eq.~(\ref{eq: eff pot total}) can be computed by using the Coleman-Weinberg formula \begin{equation}
\begin{aligned}
V_{\rm top}=-2N_c \int \frac{d^4 p}{(2\pi)^4}\log \left(1+\frac{\vert M_1^t\vert^2}{2p^2 \Pi_0^q \Pi_0^t}\frac{h^2}{\pi_Q^2}\sin^2\frac{2\pi_Q}{f}\right)\,.
\label{eq: top Coleman-Weinberg pot}
\end{aligned} \end{equation} By expanding $ V_{\rm top} $ up to $ \mathcal{O}(s_\theta^4) $ and to zero order in the fields, we obtain the effective potential \begin{equation}
\begin{aligned}
V_{\rm top}^0= \alpha_f (-s_\theta^2+s_\theta^4)+\mathcal{O}(s_\theta^6)\,,
\label{eq: top Coleman-Weinberg pot expanded}
\end{aligned} \end{equation} where \begin{equation}
\begin{aligned}
\alpha_f \equiv 4 N_c \int \frac{d^4 p}{(2\pi)^4}\frac{\vert M_1^t\vert^2}{p^2 \Pi_0^q \Pi_0^t}\,.
\label{eq: gamma f}
\end{aligned} \end{equation}

Furthermore, we obtain the potential contributions $ V_{\rm m} $ in Eq.~(\ref{eq: eff pot total}) from the hyper-fermion masses given in Eq.~(\ref{eq: Basic Lagrangian (UV)}), which is given by
\begin{equation}
\begin{aligned}  
V_{\rm m} = -2 \pi Z f^3 \text{Tr}[M_\Psi \Sigma^\dagger]+\text{h.c.}= \gamma_m c_\theta + \dots \,,
\end{aligned}
\end{equation} where we have defined $ \gamma_m\equiv 8 \pi Z f^3 m_Q   $ and $ m_Q\equiv \overline{m}_1+ \overline{m}_2$.

To leading order, the effective potential in Eq.~(\ref{eq: eff pot total}) has the form \beq
V_{\text{eff}}^0 &= -\alpha s_\theta^2 + \beta s_\theta^4 + \gamma c_\theta + \dots\,, \label{eq: leading order eff pot}
\eeq where $ \alpha = \alpha_f-\alpha_g $, $ \beta =  \alpha_f+\beta_g$ and $ \gamma = \gamma_m $. For $ \alpha \gg \beta s_\theta^2 $, we obtain the vacuum misalignment angle by minimizing the effective Higgs potential \beq
c_\theta \approx -\frac{\gamma}{2\alpha}\,. \label{eq: vacuum misalignment angle }
\eeq From the vacuum alignment and the effective potential in Eq.~(\ref{eq: eff pot total}), the Higgs mass can be expressed as \beq
m_h^2 =\frac{2s_\theta^2}{f^2}\left(\alpha+2(2-3s_\theta^2)\beta\right)\,. \label{eq: the Higgs mass}
\eeq


\section{Contributions to the oblique parameters}
\label{sec:Contributions to the oblique parameters}

Now, we are ready to collect the contributions to the oblique parameters from the modification of the Higgs couplings in Eq.~(\ref{eq: WZ masses and SM VEV}), the composite spin-1 resonances studied in Section~\ref{sec: The effective theory of the vector and axial-vector resonances} and the composite fermion resonances in Section~\ref{sec: The effective theory of the fermion resonances}. 

\subsection{Modification of the Higgs couplings}

The modification of the Higgs couplings to the EW gauge bosons in Eq.~(\ref{eq: WZ masses and SM VEV}) produces approximately the following deviations in the oblique parameters~\cite{BuarqueFranzosi:2016ooy}
\begin{equation}
\begin{aligned}
&\Delta S=\frac{1}{6\pi}\left[(1-\kappa_V^2)\ln\left(\frac{\Lambda_{\rm HC}}{m_h}\right)+\ln\left(\frac{m_h}{m_{h,ref}}\right)\right]\,,\\
&\Delta T=-\frac{3}{8\pi\cos^2\theta_W}\left[(1-\kappa_V^2)\ln\left(\frac{\Lambda_{\rm HC}}{m_h}\right)+\ln\left(\frac{m_h}{m_{h,ref}}\right)\right] 
\label{eq: mod of Higgs couplings}
\end{aligned}
\end{equation} with the Higgs couplings to the EW gauge bosons normalized to the SM ones \begin{equation}
\begin{aligned}
&\kappa_W=c_\theta + \kappa_{W,res}\simeq c_\theta +\frac{g^2}{\widetilde{g}^2}\frac{1}{2}\left(1-r^2\right)s_\theta^2\,, \\
& \kappa_Z=c_\theta + \kappa_{Z,res}\simeq c_\theta +\frac{g^2+g^{\prime 2}}{\widetilde{g}^2}\frac{1}{2}\left(1-r^2\right)s_\theta^2\,,
\label{eq: Higgs coupling mod}
\end{aligned}
\end{equation} where the correction factors  $\kappa_{W/Z,res}$ from the composite spin-1 resonances are included. For simplicity, we can neglect the term in $ g^{\prime 2} $ without significant changes, so that $ \kappa_V\sim \kappa_W \sim \kappa_Z $. Note for a typical coupling $ r\leq 1 $ these modifications contribute positively to the $ S $ parameter and negatively to the $ T $ parameter. Therefore, we need a positive contribution to $ T $ from the composite resonances to reproduce the $S,T$ correlation in the EW precision measurements obtained from the CDF and LEP measurements summarized in Eqs.~(\ref{eq: CDF results}) and~(\ref{eq: precision measurements results}).

\subsection{The composite spin-1 resonances}

At leading orders in $s_\theta$ and $g/\widetilde{g}$ the composite spin-1 contributions to the oblique parameters are given by~\cite{BuarqueFranzosi:2016ooy} \begin{equation}
\begin{aligned}
&\Delta \widehat{S}=\frac{g^2 (1-r^2)s_\theta^2}{2\widetilde{g}^2+g^2[2+(r^2-1)s_\theta^2]}, \\
&\Delta \widehat{T}=0\,, \\
& \Delta W=\frac{g^2m_W^2 [s_\theta^2 (r^2 m_V^2-m_A^2)+2m_A^2]}{m_A^2 m_V^2 (g^2[(r^2-1)s_\theta^2+2]+2\widetilde{g}^2)}\,, \\
& \Delta Y=\frac{g^{\prime 2}m_W^2 [s_\theta^2 (r^2 m_V^2-m_A^2)+2m_A^2]}{m_A^2 m_V^2 (g^{\prime 2}[(r^2-1)s_\theta^2+2]+2\widetilde{g}^2)}\,, 
\\
& \Delta X=\frac{gg^\prime s_\theta^2 m_W^2(m_A^2-r^2m_V^2)}{m_A^2 m_V^2 \sqrt{(g^2[(r^2-1)s_\theta^2+2]+2\widetilde{g}^2)(g^{\prime 2}[(r^2-1)s_\theta^2+2]+2\widetilde{g}^2)}}\,.
\label{eq: vector axial-vector cont}
\end{aligned}
\end{equation} In our analysis, we are going to use the notation adopted by the PDG and rescale \begin{equation}
\frac{\alpha_Z S}{4s_W^2}=\Delta\widehat{S}-\Delta Y-\Delta W\,, \quad \quad \quad \alpha_Z T = \Delta\widehat{T}-\frac{s_W^2}{1-s_W^2}\Delta Y\,,
\label{eq:}
\end{equation} where the Weinberg angle $ s_W\equiv \sin\theta_W \simeq 0.223 $ and the fine-structure constant at the $ Z $ boson mass $ \alpha_Z\equiv \alpha(m_Z)\simeq 1/127.916 $. Again these resonances contribute positively to $ S $ and a negligible (negative) $ T $. Therefore, the positive $ T $ contribution must come from the fermion resonances.

\begin{figure}[t]
	\centering
	\includegraphics[width=1\textwidth]{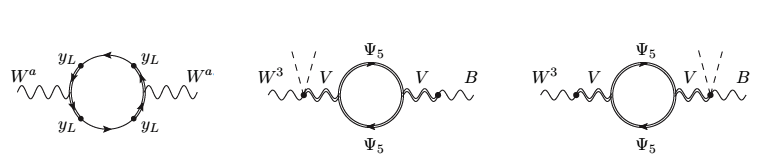}
	\caption{Radiative contributions to the $ T $ parameter (left panel diagram) and the $ S $ parameter (middle and right panel diagrams). The straight and wavy double lines represent, respectively, the top-partner spurions and the composite spin-1 resonances. }
	\label{fig: diagrams}
\end{figure}

\subsection{The fermion resonances}

We calculate the contribution from the fermion resonances in three different scenarios, where we include the contributions from \begin{itemize}
	\item[(i)] only a light top-partner singlet, $ T_1 $, 
	\item[(ii)] only a light top-partner from the fiveplet, $ T $, and 
	\item[(iii)] both the singlet and fiveplet.
\end{itemize}

\textbf{Scenario (i):} In this case in which only the top-partner singlet, $ T_1 $, is light, the $ T $ parameter receives a sizable positive contribution from loops of the fermion resonances as the diagram in the left-handed panel of Figure~\ref{fig: diagrams}, while the contribution to the $ S $ parameter of the two other diagrams to the right is negligible. At leading order in $ s_\theta $, the explicit results read\begin{equation}\begin{aligned}
&\Delta \widehat{S}\simeq 0\,, \\
&\Delta \widehat{T}=\frac{3}{64\pi^2}\frac{y_{L1}^4 m_1^4 v_{\rm SM}^2}{(m_1^2+y_{R1}^2f^2)^3}\left[1+\frac{2y_{R1}^2f^2}{m_1^2}\left(\log\left(\frac{2(m_1^2+y_{R1}^2 f^2)^2}{y_{L1}^2 y_{R1}^2 f^4 s_\theta^2}\right)-1\right)\right]\,.
\label{eq: singlet S and T cont}
\end{aligned} 
\end{equation} In a large part of the parameter space, this contribution can compensate the
negative $ T $ contribution from the modification of the Higgs couplings in Eq.~(\ref{eq: mod of Higgs couplings}) and reproduce the $S,T$ correlations found in fits to observations.

\begin{figure}[t]
	\centering
	\includegraphics[width=0.385\textwidth]{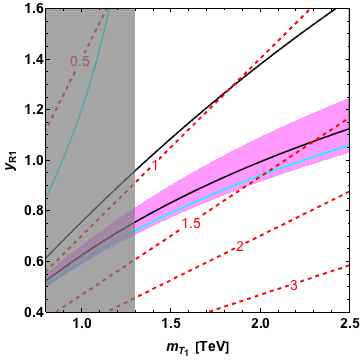}
	\includegraphics[width=0.6\textwidth]{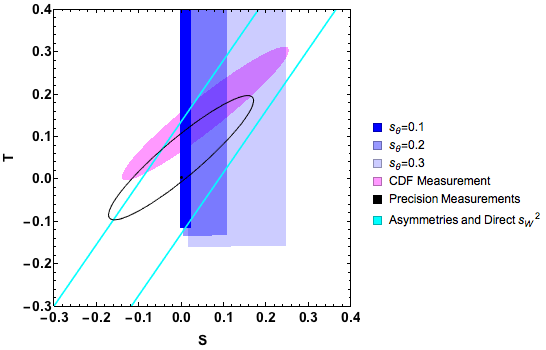}
	\caption{Comparison of the new CDF measurement of the W mass (magenta band) to the parameter space of the $ \SU(4)_Q/\text{Sp}(4)_Q $ Composite Higgs model, including only the $ S $ and $ T $ contributions from the fermion resonance singlet. Between the black lines or in the ellipse the precision measurements are fulfilled, while the cyan lines indicate the bound stemming from asymmetries and direct $ s_W^2 $ determination. All bounds are shown at $ 2\sigma $ confidence level. \textit{Left panel}: The bands in the $ m_{T_1} $--$ y_{R1} $ parameter space for $ s_\theta =0.2 $, $ m_{V,A}=3 $~TeV, $ \widetilde{g}=3 $ and $ r=0.8 $, where the dashed red lines represent the values of $ y_{L1} $. The gray region is excluded by the lower bound on $ m_{T_1}\gtrsim 1.3 $~TeV from CMS and ATLAS~\cite{ ATLAS:2018ziw,CMS:2018zkf,CMS:2017ynm,CMS:2019eqb}. \textit{Right panel}: The bands in the $ S $--$ T $ parameter space, where the blue regions represent the parameter space of the model for $ s_\theta=0.1,0.2,0.3 $. We have scaned over $ \widetilde{g} \geq 3 $, $ 0.5 \leq y_{R1} \leq 5 $, $ m_{V,A}\geq 3 $~TeV and $ 0 \leq r \leq 1 $, while $ m_{T_1} \geq 1.3 $~TeV from CMS and ATLAS~\cite{ ATLAS:2018ziw,CMS:2018zkf,CMS:2017ynm,CMS:2019eqb}. }
	\label{fig: only singlet}
\end{figure} 

In the present model example, we have nine free parameters $ y_{L1,R1} $, $ m_{1} $, $ m_{V,A} $, $ \widetilde{g} $, $ r $, $ f $ and $ s_\theta $ in the above expressions of the $ S $ and $ T $ parameters. In the following, we fix the parameters of the composite spin-1 resonances $ m_{V,A}=3 $~TeV, $ \widetilde{g}=3 $ and $ r=0.8 $. Changes in these parameters have small effects on the following conclusions. Furthermore, we can determine $ y_{L1} $ by fixing the top mass for $ m_{T}\gg m_{T_1} $ in Eq.~(\ref{eq: top mass approx}), determine $ f $ by fixing the EW VEV in Eq.~(\ref{SM VEV}) and replace $ m_1 $ with the lightest top-partner mass $ m_{T_1} $ in Eq.~(\ref{eq: fermion resonance masses}). This reduces the number of free parameters to three: $ y_{R1} $, $ m_{T_1} $ and the vacuum alingment angle $ s_\theta $. 

In Figure~\ref{fig: only singlet}, the parameter space of the minimal CH model with only the fermion resonance singlet is shown which is constrained by the new CDF measurement in Eq.~(\ref{eq: CDF results}), depicted by the magenta band or ellipse. Between the black lines or inside the black ellipse the precision measurements in Eq.~(\ref{eq: precision measurements results}) are satisfied, while the cyan lines indicate the bound stemming from asymmetries and direct $ s_W^2 $ determination in the SM~\cite{Haller:2018nnx}. All bounds are shown at ($ 2\sigma $) $ 95\% $ confidence level. In the left-handed panel, the bands are shown in the $ m_{T_1} $--$ y_{R1} $ parameter space with $ s_\theta =0.2 $, where the dashed red lines represent the values of $ y_{L1} $. The lower bound on $ m_{T_1}\gtrsim 1.3 $~TeV is given by CMS and ATLAS~\cite{ ATLAS:2018ziw,CMS:2018zkf,CMS:2017ynm,CMS:2019eqb}, which excludes the gray region. There is thus a viable area in the parameter space with $ y_{L1,R1} $ of order unity that satisfies the new CDF and/or LEP measurements along with the lower bounds on $ m_{T_1} $ and from the asymmetries and direct $ s_W^2 $ determination. In the right-handed panel, the bands are shown in the $ S $--$ T $ parameter space, where the blue regions represent the parameter space of the model for $ s_\theta=0.1,0.2,0.3 $. We have scanned over $ \widetilde{g} \geq 3 $, $ 0.5 \leq y_{R1} \leq 5 $, $ m_{V,A}\geq 3 $~TeV and $ 0 \leq r \leq 1 $, while $ m_{T_1} \geq 1.3 $~TeV from CMS and ATLAS~\cite{ ATLAS:2018ziw,CMS:2018zkf,CMS:2017ynm,CMS:2019eqb}. Thus, there are viable regions in the parameter space for $ s_\theta\leq 0.3 $. Therefore, the constraints on $ s_\theta $ from the EW precision measurements (also the recent CDF results) are weaker than the bound from the Higgs couplings to the EW gauge bosons in Eq.~(\ref{eq: WZ masses and SM VEV}), setting the constraint $ s_\theta \lesssim 0.3 $~\cite{deBlas:2018tjm}. \\

\begin{figure}[t]
	\centering
	\includegraphics[width=0.9\textwidth]{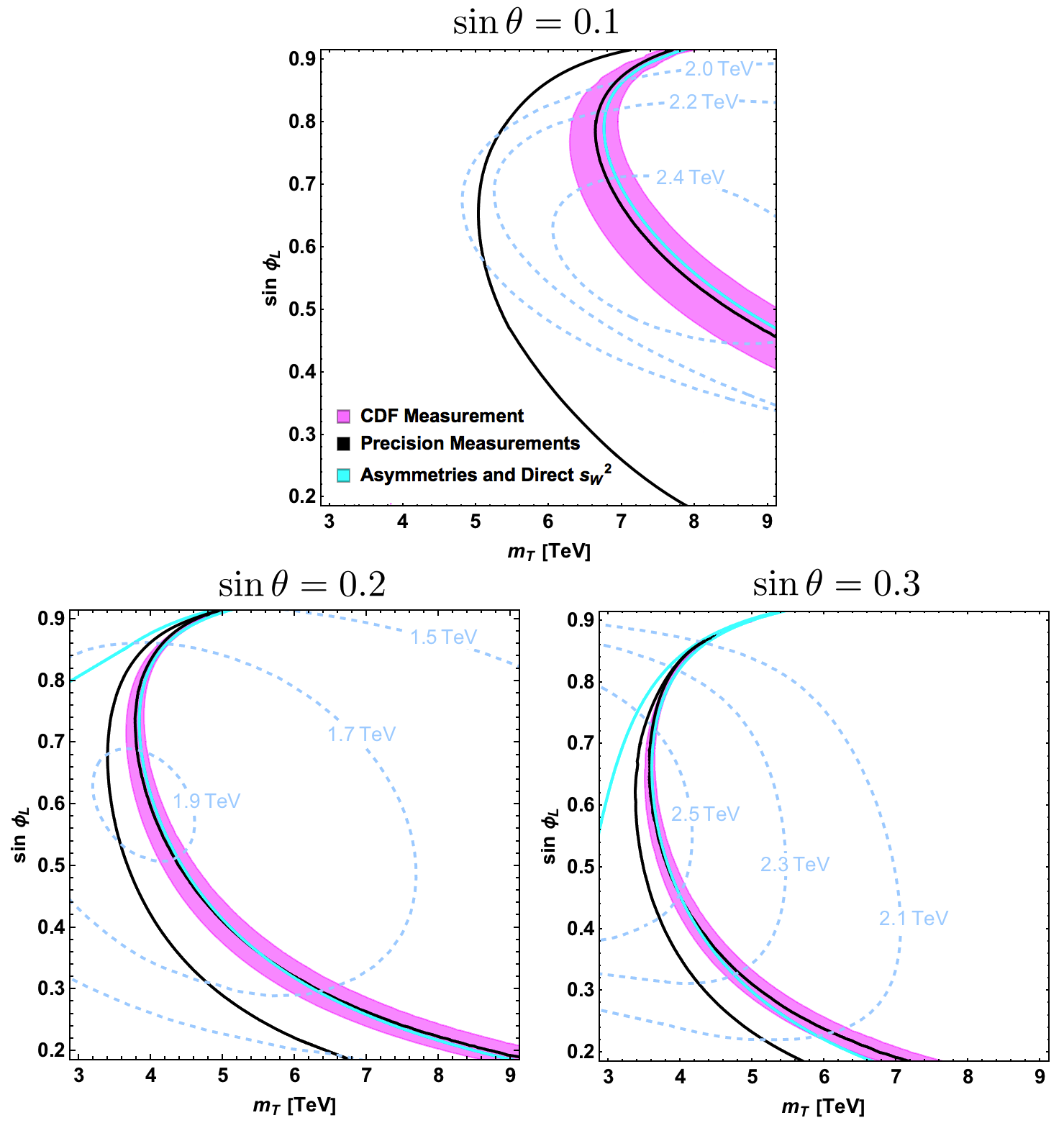}
	\caption{Comparison of the new CDF measurement of the W mass (magenta band) to the $ m_T $--$ \sin \phi_L $ parameter space of the minimal CH model for $ s_\theta =0.1,0.2,0.3 $, including the $ S $ and $ T $ contributions from both the fermion resonance singlet and fiveplet. Between the black lines the precision measurements are fulfilled, while the cyan lines indicate the bound stemming from asymmetries and direct $ s_W^2 $ determination. All bounds are shown at $ 2\sigma $ confidence level and for $ m_{V,A}=3 $~TeV, $ \widetilde{g}=3 $ and $ r=0.8 $. 
The dashed lines represent the top-partner singlet mass, $ m_{T_1} $. }
	\label{fig: both singlet and fiveplet}
\end{figure}   
  
\textbf{Scenario (ii):} Now only a light top-partner, $ T $, from the fiveplet of the fermion resonance spectrum is included. In this case, the dominant $ S $ contribution (the two diagrams to the right in Figure~\ref{fig: diagrams}) comes from loops of fermion resonances, where the lightest vector resonances mix with the EW gauge bosons in Eq.~(\ref{eq: VA lag}). This contribution is given by the logarithmical corrections \begin{equation}
\Delta \widehat{S}=\frac{g^2s_\theta^2}{8\pi^2}(1-c_L^2-c_R^2)\log\left(\frac{m_V^2}{m_5^2}\right)\,.
\label{eq: S cont fiveplet}
\end{equation} From now on, we choose $ c_L=c_R=0 $ in Eq.~(\ref{eq: comp lagrangian}) by assuming that these interactions are weak. Moreover, the $ T $ contribution from the diagram in the left panel in Figure~\ref{fig: diagrams} with the fiveplet top-partner, $ T $, running in the loop is given by \begin{equation}
\Delta \widehat{T}=-\frac{y_{L5}^4}{32\pi^2}\left(\frac{ v_{\rm SM}}{m_5}\right)^2\,,
\label{eq: T for scenario ii}
\end{equation} which is always negative. Thus, this scenario has difficulty satisfying the EW precision measurements. \\

\textbf{Scenario (iii):} In the last scenario, we include the contributions from both the singlet and fiveplet, where all the $ S $ and $ T $ contributions in Eqs.~(\ref{eq: mod of Higgs couplings})-(\ref{eq: T for scenario ii}) must be included in the following calculations. Therefore, the positive $ T $ contribution from the top-partner singlet in Eq.~(\ref{eq: singlet S and T cont}) has to compensate the negative contributions from the modification of the Higgs couplings in Eq.~(\ref{eq: mod of Higgs couplings}) and the fiveplet in Eq.~(\ref{eq: T for scenario ii}). 

As in the case (i), we fix $ m_{V,A}=3 $~TeV, $ \widetilde{g}=3 $ and $ r=0.8 $ due to the fact that changes of them have small effects on the following results. We assume $ y_{L1}=y_{L5}\equiv y_L $ and $ y_{R1}=y_{R5}\equiv y_R $, while we use the expressions of the EW VEV in Eq.~(\ref{SM VEV}), the top mass in Eq.~(\ref{eq: top mass}), the masses of the two lightest fermion resonances $ T_{1} $ and $ T $ in Eq.~(\ref{eq: fermion resonance masses}) and the Higgs mass in Eq.~(\ref{eq: the Higgs mass}). These expressions reduce the number of free parameters from ten $ y_{L1,R1,L5,R5} $, $ m_{T_1,T} $, $ m_{1,5} $, $ f $ and $ s_\theta $ to three $ m_{T} $, $ s_{\phi_L} $ and $ s_\theta $, where we have defined the $ q_L $ compositeness angle given by\begin{equation}\begin{aligned}
\sin \phi_L\equiv \frac{y_L f}{\sqrt{m_5^2+y_L^2 f^2}}\,.
\label{eq:}
\end{aligned} 
\end{equation} 

In Figure~\ref{fig: both singlet and fiveplet}, the $ m_{T} $--$ s_{\phi_L} $ parameter space is shown for the minimal CH model with both the fermion resonance singlet and fiveplet, which is constrained by the new CDF measurement of the W mass, depicted by the magenta band. The same bands as in Figure~\ref{fig: only singlet} are depicted and all at ($ 2\sigma $) $ 95\% $ confidence level. In this figure, we consider these bands for $ s_\theta =0.1,0.2,0.3 $. The dashed lines represent the values of the top-partner singlet mass, $ m_{T_1} $. As in scenanio (i), the bound on $ s_\theta $ from the Higgs couplings to the EW gauge bosons in Eq.~(\ref{eq: WZ masses and SM VEV}), setting the constraint $ s_\theta \lesssim 0.3 $~\cite{deBlas:2018tjm}, is stronger than  the constraints from both the recent CDF and the LEP precision measurements. Furthermore, the lower bound on the lightest top-partner (the singlet) $ m_{T_1}\gtrsim 1.3 $~TeV from CMS and ATLAS~\cite{ ATLAS:2018ziw,CMS:2018zkf,CMS:2017ynm,CMS:2019eqb} does not narrow the parameter space shown in Figure~\ref{fig: both singlet and fiveplet} for these values of $ s_\theta $, only for large $ m_{T} $ as in case (i). For small $ s_{\phi_L} $, this lower bound on $ m_{T_1} $ leads to the constraint $ m_T\lesssim 77 $~TeV ($ s_{\phi_L} \gtrsim 0.019 $) for $ s_\theta=0.2 $ and no constraints for $ s_\theta=0.1 $ and $ 0.3 $. However, for $ s_\theta=0.3 $, there is the upper bound $ m_T\lesssim 26 $~TeV from the condition $ s_{\phi_L} > 0 $. Finally, we will not consider the bounds in the case with large $ s_{\phi_L} $ due to the fact that the bands depicted in the figure are very narrow in these regions and therefore very fine-tuned in $ s_{\phi_L} $. 

\section{Generalization to other 4D CH models}

The EW precision tests of the minimal CH model studied in this paper can be generalized to non-minimal models. In general, CH model arising from an  underlying four-dimensional gauge theory, contain at least one HC representation $\mathcal{R}$ with $N_f$ Weyl hyper-fermions, providing a SM-like composite pNGB Higgs multiplet with custodial symmetry after the condensation. Depending on this representation, there exist three possible types of vacuum cosets~\cite{Peskin:1980gc}: $ \SU(N_f)/\SO(N_f) $ for real $\mathcal{R}$, $ \SU(N_f)/\SP(N_f) $ for pseudo-real $\mathcal{R}$ and $ \SU(N_f)\otimes \SU(N_f)\otimes \UU(1)/ \SU(N_f)\otimes \UU(1)$ for complex $\mathcal{R}$. For these three types, the minimal CH cosets contain $N=5$ in the real case~\cite{Dugan:1984hq}, $N=4$ in both the pseudo-real~\cite{Galloway:2010bp} and the complex cases~\cite{Ma:2015gra}. In terms of pNGB spectrum, the pseudo-real case considered in this paper has the fewest number of pNGB states with only five. 



For all those CH models, the Higgs couplings to the EW gauge bosons are generally modified by $ c_\theta $ as in Eq.~(\ref{eq: Higgs coupling mod}), providing the $ S $ and $ T $ contributions in Eq.~(\ref{eq: mod of Higgs couplings}). The contributions from the spin-1 resonances in Eq.~(\ref{eq: vector axial-vector cont}) are in general determined by the lightest vector and axial-vector $\SU(2)_V$ triplets so also this contribution will only vary slightly between models as long as the underlying fermion dynamics is not vastly different (e.g. QCD-like versus near-conformal). In all cases, variations will have a modest impact due to the smallness of the $S$ parameter and especially $T$ parameter contributions. Finally, the top-partner singlet, similar to the singlet $ T_1 $, are also present in the non-minimal CH models, where either the left- or right-handed top quarks are transforming as two-index anti-symmetric under the unbroken chiral symmetry group for both the coset $ \SU(N_f)/\SP(N_f) $ and $ \SU(N_f)\otimes \SU(N_f)\otimes \UU(1)/ \SU(N_f)\otimes \UU(1)$ or transforming two-index symmetric for the coset $ \SU(N_f)/\SO(N_f) $ (model examples of these kinds are considered in Ref.~\cite{Ferretti:2016upr} with the minimal cosets $ \SU(5)/\SO(5) $, $ \SU(4)/\SP(4) $ and $ \SU(4)^2 \otimes \UU(1)/\SU(4)\otimes \UU(1) $). The presence of the $ T_1 $ type singlets will contribute positively to the $ T $ parameter and provided this is the lightest fermion resonance this contribution will dominate leading to the $S,T$ parameter correlation and offset observed in current precision electroweak measurements and underscored by the recent CDF W-mass measurement. On the other hand, the higher dimensional multiplets, for example the fiveplet considered in this paper, may contribute with negative $ S $ and $ T $ contributions with similar form as given by Eqs.~(\ref{eq: S cont fiveplet}) and~(\ref{eq: T for scenario ii}), respectively, which can be suppressed by their masses. Furthermore, we can not completely exclude the possibility that those multiplets can contribute positively to the oblique parameters. 

We leave a detailed study of the contributions from those multiplets for future work. Nevertheless, it is likely that we can draw similar conclusions for these non-minimal models as in the case of the minimal model example. 

\section{CH models with coset $ \SO(N_f)/\SO(N_f-1) $}

\begin{figure}[t]
	\centering
	\includegraphics[width=0.9\textwidth]{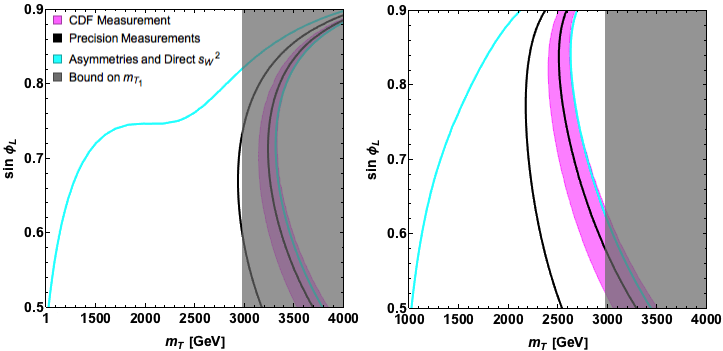}
	\caption{Comparison of the new CDF measurement of the W mass (magenta band) to the $ m_T $--$ \sin \phi_L $ parameter space of the $ \SO(5)/\SO(4) $ CH model for $ s_\theta =0.2$, including the $ S $ and $ T $ contributions from both the fermion resonance singlet and fourplet. Between the black lines the precision measurements are fulfilled, while the cyan lines indicate the bound stemming from asymmetries and direct $ s_W^2 $ determination. All bounds are shown at $ 2\sigma $ confidence level and for $ m_{V,A}=3 $~TeV, $ \widetilde{g}=3 $ and $ r=0.8 $. The coupling $ y_R $ is determined up to a twofold ambiguity, where the bands in the $ m_{T} $--$ \sin \phi_L $ parameter space are shown for the smallest (largest) $ y_R $ solution in the left (right) panel. The gray regions are excluded by the lower bound on $ m_{T_1}\gtrsim 1.3 $~TeV from CMS and ATLAS~\cite{ ATLAS:2018ziw,CMS:2018zkf,CMS:2017ynm,CMS:2019eqb}. 
	}
	\label{fig: both singlet and fourplet}
\end{figure} 

Notice that the top-partner resonance $ T_5 $ in the fiveplet in Eq.~(\ref{eq: the fiveplet}) does not contribute to the oblique parameters due to the fact that it does not mix with the top quark in the mass matrix in Eq.~(\ref{eq: top mass matrix}). In a CH model with coset $ \SO(5)/\SO(4) $, proposed in Ref.~\cite{Agashe:2004rs}, the fiveplet is replaced by a fourplet without the top-partner $ T_5 $ and, therefore, the results in this paper will also be valid for this CH model. However, this model is difficult to realize in a 4D theory. For example, such a setup may require a 5D theory, described by the holographic approach~\cite{Agashe:2004rs}. Generally, in models with cosets $ \SO(N_f)/\SO(N_f-1) $ (e.g. $ \SO(5)/\SO(4) $), the explicit masses of the hyper-fermions in Eq.~(\ref{eq: Basic Lagrangian (UV)}) can not be included, leading to a modification of the Higgs mass in Eq.~(\ref{eq: the Higgs mass}): \beq
m_h^2 =\frac{8\beta}{f^2} s_\theta^2 c_\theta^2 \,, \label{eq: the Higgs mass SO(5)/SO(4)}
\eeq where $ s_\theta^2=\alpha/(2\beta) $ by minimizing the Higgs potential in Eq.~(\ref{eq: leading order eff pot}) with $ \gamma_m =0 $. In Ref.~\cite{Panico:2015jxa}, a relation between the masses of the Higgs boson, the top quark and the top partners is derived: \begin{equation}
\begin{aligned}
\frac{m_h^2}{m_t^2}\simeq \frac{2N_c}{\pi^2}\frac{m_{T_1}^2m_{T}^2}{f^2}\frac{\log (m_{T}/m_{T_1})}{m_{T}^2-m_{T_1}^2}\,,
\label{eq: Higgs top mass relation}
\end{aligned}
\end{equation} which relates the masses of the singlet  and fourplet top-partners, $ m_{T_1} $ and $ m_{T} $, by fixing the Higgs and top mass. By repeating the calculations of the oblique parameters from the modification of the Higgs couplings and the spin-1 resonances, we obtain similar expressions of them as in Eqs.~(\ref{eq: mod of Higgs couplings}) and~(\ref{eq: vector axial-vector cont}), respectively.

In the following, we choose $ y_{L1}=y_{L5}\equiv y_L $ and $ y_{R1}=y_{R5}\equiv y_R $ for simplicity. For $ s_\theta =0.2 $, $ m_{V,A}=3 $~TeV, $ \widetilde{g}=3 $ and $ r=0.8 $, the $ m_{T} $--$ \sin\phi_L $ parameter space is shown in Figure~\ref{fig: both singlet and fourplet} for the CH model with coset $ \SO(5)/\SO(4) $, where both the fermion resonance singlet and fourplet are included. This parameter space is constrained by the same bands as in Figure~\ref{fig: both singlet and fiveplet}, while the gray regions are excluded by the lower bound on $ m_{T_1}\gtrsim 1.3 $~TeV from CMS and ATLAS~\cite{ ATLAS:2018ziw,CMS:2018zkf,CMS:2017ynm,CMS:2019eqb}. The coupling $ y_R $ is determined up to a twofold ambiguity, where the bands are shown for the smallest and largest $ y_R $ solution in the left and right panel in the figure, respectively. This analysis can also be generalized to $ \SO(N_f)/\SO(N_f-1) $ by following the same arguments as presented in the previous section. In the case where either the left- or right-handed top quarks are transforming two-index antisymmetric under the unbroken $ \SO(N_f-1) $, there will always be a singlet top-partner, $ T_1 $, contributing positively to the $ T $ parameter. Moreover, the same top-partner, $ T $, of the fourplet will always be present in one of the larger multiplets. By assuming that the additional top-partners are heavier than $ T_1 $ and $ T $, the conclusions from the analysis of the cosets $ \SO(N_f)/\SO(N_f-1) $ will be almost unchanged compared to the above analysis of the minimal $ \SO(5)/\SO(4) $ coset.     

\section{Conclusions}
\label{sec:Conclusions}
In this paper, we have computed the oblique electroweak $S$ and $T$ parameters in a minimal Composite Higgs model arising from an underlying 4D strongly interacting fermion-gauge theory. The CH sector consists of four Weyl fermions in a pseudo-real representation of a new strongly interacting gauge group and the SM fermion masses arise from linear mixings of interaction eigenstate SM fermions with composite fermions arising from additional strongly interacting fermions of the model. The obtained results generalize to other non-minimal CH models arising from an underlying 4D gauge-fermion theory with fermion partial compositeness.

The Composite Higgs sector leads to a small and positive electroweak $S$ parameter of $O(1/6\pi)$ along with a negative $T$ parameter of similar size. The dominant source for these are the coposite Higgs couplings to SM gauge bosons which are reduced with respect to the SM Higgs. The contributions from spin-1 resonances are smaller still, so the Composite Higgs sector in isolation is not in agreement with the best fit $S$ and $T$ parameter values which favor a positive correlation between $S$ and $T$. However, the fermion partial compositeness sector leads to a positive $T$ parameter in a large part of parameter space for which we find good agreement between the model and current electroweak precision measurement of the $S$ and $T$ parameters, even including the recent CDF $W$-boson mass measureent, including the offset and correlation of $S,T$ with respect to the SM predictions. This requires light composite fermions around the TeV scale, but in agreement with current LHC direct search limits from ATLAS and CMS. 
If the 2022 CDF results are interpreted in terms of Composite Higgs models, they may offer a striking first window into origin of the SM fermion masses. It will be interesting to further explore the implications for other possible fermion mass origins, in particular ``Extended Technicolor'' (ETC)~\cite{Dimopoulos:1979es} and partially composite Higgs models \cite{Alanne:2017rrs}, as well as the complementarity of precision electroweak meaurements with other probes of the SM fermion mass origin \cite{Alanne:2016rpe}.


\section*{Acknowledgements}
MTF and MR acknowledge partial funding from The Independent Research Fund Denmark, grant numbers DFF 6108-00623 and DFF 1056-00027B, respectively. We would also like to thank Giacomo Cacciapaglia for useful discussions.

\appendix

\end{document}